\documentclass[journal]{IEEEtran}

\usepackage{cite}      

\usepackage{graphicx}  

\usepackage{subfigure} 

\usepackage{stfloats}  

\usepackage{amsmath}   
\begin{document}

%
\title{Radiation Tolerance of Fully-Depleted P-Channel CCDs Designed for the SNAP Satellite}
%
%
\author{Kyle~Dawson, Chris~Bebek, John~Emes, Steve~Holland,
        Sharon~Jelinsky, Armin~Karcher,
	William~Kolbe, Nick~Palaio, Natalie~Roe, Juhi~Saha, Koki~Takasaki, and Guobin~Wang %
\thanks{The authors are with Lawrence Berkeley National Laboratory,
	One Cyclotron Rd,
	Berkeley, CA 94720} %
\thanks{Further author information:  e-mail kdawson@lbl.gov}
}

\maketitle

\begin{abstract}
Thick, fully depleted p-channel charge-coupled devices (CCDs) have been developed at the
Lawrence Berkeley National Laboratory (LBNL).
These CCDs have several advantages over conventional thin, n-channel CCDs, including enhanced quantum efficiency
and reduced fringing at near-infrared wavelengths 
and improved radiation tolerance.
Here we report results from the irradiation of CCDs with 12.5 and 55~MeV
protons at the LBNL 88-Inch Cyclotron and with 0.1 - 1~MeV electrons
at the LBNL $^{60}$Co source.
These studies indicate that the LBNL CCDs perform well
after irradiation, even in the parameters in which significant degradation is observed in other CCDs: charge transfer
efficiency, dark current, and isolated hot pixels.
Modeling the radiation exposure over a six-year mission lifetime with no annealing,
we expect an increase in dark current of 20~e$^-$/pixel/hr, and a degradation of 
charge transfer efficiency in the parallel direction of $3 \times 10^{-6}$  and
$1 \times 10^{-6}$ in the serial direction.
The dark current is observed to improve with an annealing cycle,
while the parallel CTE is relatively unaffected and the serial CTE
is somewhat degraded.
As expected, the radiation tolerance of the p-channel LBNL CCDs is
significantly improved over the conventional n-channel CCDs that are currently employed in space-based telescopes such as the
Hubble Space Telescope.
\end{abstract}

\begin{keywords}
Astrophysics and Space Instrumentation, Radiation Damage Effects
\end{keywords}

\section{Introduction}
The SuperNova/Acceleration Probe (SNAP) is a proposed space-based telescope
dedicated to the study of dark energy through the observations of Type Ia supernovae (Ia SNe)
and a deep, wide area weak lensing survey\cite{greg}.  
From its orbit at the second Earth-Sun Lagrange point (L2), SNAP will carry out two surveys:
a deep survey of $7.5$ square degree field with 4-day cadence repeat visits
over a period of 22 months to 
discover and obtain light curves and spectra of over 2000 Ia SNe in the 
redshift range $0.3 < z < 1.7$; and a wide area
weak lensing map to study the growth of large scale structure that will cover
1000 square degrees per year to a depth of AB magnitude $28.0$ in the optical filters. 
In an extended 6 year SNAP mission, the weak lensing survey covers
4000 square degrees and the mission lifetime.

The telescope is designed with a $0.7$ square
degree instrumented field of view divided evenly between 36 CCDs and 36 HgCdTe detectors.
The SNAP observing strategy implements a four-point dither pattern with an exposure time of 300 seconds
to recover spatial information from the undersampled optics and to reject cosmic rays.
The focal plane will be passively cooled to 140~K.  Nine fixed filters cover
the wavelength range 400~nm to 1700~nm.
With a diffraction limited point spread function (PSF) of $0.1$ arcseconds at~800 nm and zodiacal-dominated
background, SNAP will have significantly improved resolution
and decreased contamination from sky background compared to ground based telescopes.

The SNAP focal plane design uses thick, fully depleted CCDs developed at LBNL~\cite{steve1,steve} for visible to
near IR observations in six bandpass filters.
In space, these detectors will be exposed to significant radiation, primarily 
solar protons.
In this paper we investigate the effects of six years of radiation at L2
on SNAP CCDs in order to qualify them
for use in a space mission.
In \S\ref{section:CCD} we describe the SNAP CCDs and the specifications for performance.
The space environment and expected radiation exposure are discussed in \S\ref{section:space}.
Irradiation using the 88-Inch Cyclotron and the $^{60}$Co source
at LBNL is described in \S\ref{section:irrad} and \S\ref{section:Co60} respectively. 
CCD performance after proton irradiation
is reported in \S\ref{section:results} and after $^{60}$Co irradiation in \S\ref{section:Co60 results}.
Finally, we present an interpretation of the results in the context of the SNAP mission in
\S\ref{section:discussion} and the conclusions in \S\ref{section:conclusion}.

\section{CCD Requirements} \label{section:CCD}

\begin{table*}[t!]
\renewcommand{\arraystretch}{1.0}
\caption{Specifications for SNAP CCDs}
\label{table:specs}
\centering
\begin{tabular}{|c||c||c|}
Quantity & Requirement & Achieved (pre-irradiation) \\
\hline
Wavelength Coverage & $400-1000$ nm & $400-1000$ nm \\
Quantum & $>80\%$ at $600-950$ nm & $>80\%$ at $600-950$ nm \\
Efficiency & $>50\%$ at 1000 nm & $>50\%$ at 1000 nm\\
Readout Time & 30 seconds & 30 seconds \\
Read Noise & 6 e$^-$ & 4 e$^-$\\
Diffusion (RMS)  & $6\mu$m & $4\mu$m\\
Defect Pixels$^a$ & To Be Determined & $<0.1\%$\\
Dark Current$^a$ & 100 e$^-$/hr & $3-4$ e$^-$/hr \\
Serial CTE$^a$ & To Be Determined & 0.999 999\\
Parallel CTE$^a$ & To Be Determined & 0.999 999\\
\hline
\multicolumn{3}{|l|} {$^a$expected to deteriorate with irradiation} \\
\end{tabular}
\end{table*}

SNAP CCDs have been designed for back-illumination on 200~$\mu$m thick,
fully-depleted, high-resistivity silicon.
A factor of ten increase in thickness over conventional CCDs provides
vastly improved sensitivity toward wavelengths of $1 \mu$m and negligible fringing
effects caused by multiple reflections in the silicon\cite{QE1,QE2}.
The CCDs are depleted through application of a substrate bias voltage across the full thickness.
The spatial resolution can be improved by increasing the bias voltage up to 200~V\cite{steve},
with a nominal operating voltage of 100~V for the SNAP mission.
The SNAP focal plane will be populated with 36 LBNL CCDs, each having
$3512 \times 3512$ $10.5 \, \mu$m pixels.

The objectives of the SNAP experiment are to extract point-source SNe from
diffuse host galaxies and to resolve distant galaxies for weak lensing studies.
The specifications for CCD performance are therefore governed by requirements for
preservation of the point spread function (PSF), high quantum efficiency (QE), charge transfer efficiency (CTE)
and signal-to-noise ratio.
In Table \ref{table:specs} we list the specifications for the SNAP CCDs.
As can been seen in the table, each of these requirements has been met in
the current design of SNAP style devices before radiation exposure.

CCD performance is expected to degrade in a radiation environment
due to bulk damage from non-ionizing energy loss (NIEL)
and due to charging of oxide layers from ionizing radiation.
The major bulk damage in conventional n-channel CCDs is caused by traps generated in the
formation of phosphorus-vacancy centers\cite{pvvac}.
This bulk damage manifests itself through decreased charge transfer efficiency,
increased dark current, and isolated
hot pixels.
The LBNL p-channel CCDs are fabricated on high-resistivity n-type silicon
with boron implanted channels.
In the p-channel CCDs, divacancy states are expected to
be the dominant hole trap \cite{spratt05,spratt,hopkinson}.
It has been predicted that divacancy formation in p-channel CCDs
is less favorable than phosphorus-vacancy traps in n-channel CCDs \cite{spratt},
and prior studies have shown improved performance after  radiation exposure \cite{spratt05,marshall,bebek}.

Ionizing radiation is expected to result in charging of oxide layers,
requiring adjustment of pixel gate voltages and output source follower transistor biasing.
Significant increases in dark current after ionizing radiation have also been observed in p-channel
CCDs\cite{spratt}.
In this work we investigate the effects of both kinds of radiation damage on SNAP CCDs,
focusing on generation of dark current, hot pixels, and decrease in charge transfer
efficiency.

\section{Space Environment and Expected Dose} \label{section:space}

The SNAP satellite will 
orbit at the L2 Lagrange point, approximately $1.5 \times 10^6$~km from Earth.
At this distance, solar protons dominate the total radiation dose.
To estimate the total exposure at L2, we use the  Emission of Solar Protons (ESP) model
described in \cite{solar}
and the Space Environment Information System (SPENVIS)\cite{spenvis}.
In SPENVIS, the solar model is simplified as a cycle with seven years at maximum activity with constant
exposure and four years at minimum activity with no exposure.
The model provides a statistical estimate of the fluence as a function of confidence interval based on data
from the past three solar cycles.
A simple shielding model is used in which a spherical aluminum shell surrounds the detectors.
The propagation of particles through the shielding is also simplified;
showers and secondary particles are not modeled.   With these simplifications, we make a first-order estimate of the effects of radiation on the SNAP 
visible detectors.  A more
detailed Monte Carlo simulation of the propagation of particles through the 
structures of the satellite will be performed at a later date.

\begin{figure}[!h]
\centering
\includegraphics[width=3.45in]{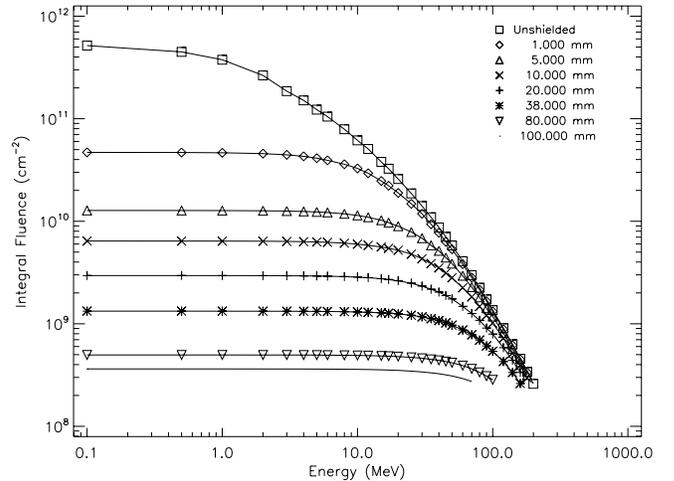}
\caption{Spectrum of incident particles for various shielding thicknesses (Al equivalent).
Results indicate $95\%$ upper limits assuming a six year mission with launch date January 1, 2014.
A shielding thickness of $\sim 38$~mm is the average amount of shielding of the SNAP focal plane.}
\label{fig:spectrum}
\end{figure}

\begin{figure}[!h]
\centering
\includegraphics[width=3.45in]{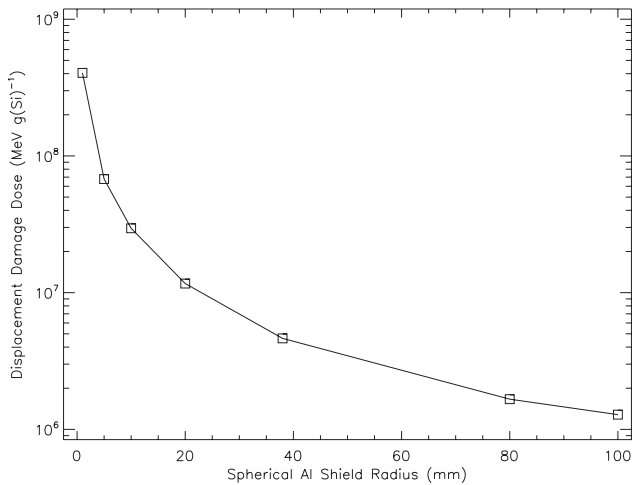}
\caption{NIEL dose as a function of shielding thickness.
Results indicate $95\%$ upper limits assuming a six year mission with launch date January 1, 2014.
For comparison, the average amount of shielding surrounding the SNAP focal plane is equivalent
to 47 mm of aluminum.}
\label{fig:niel}
\end{figure}

Assuming a six year extended mission with a January 1, 2014 launch date,
we estimate the accumulated radiation exposure for the SNAP CCDs at the $95\%$
confidence level.
Figure~\ref{fig:spectrum} shows the spectrum of protons incident on the
detectors for various shield thicknesses predicted by ESP and SPENVIS. 
Similarly, Figure~\ref{fig:niel} reports the integrated non-ionizing energy loss (NIEL) 
as a function of shield thickness.

\begin{table*}[!b]
\caption{List of irradiated CCDs.  BI refers to back-illuminated devices while FI refers to front-illuminated devices. }
\label{table:devices}
\centering
\begin{tabular}{|c||c||c||c||c||c|}
Device $\#$ & Format & Radiation & Energy & Warm/Cold & Dose \\
\hline
1 & $3512 \times 3512$ pixels, FI & proton & 55 MeV & warm & $5 \times 10^9$, $1 \times 10^{10}$, $5 \times 10^{10}$, $1 \times 10^{11}$ protons/cm$^2$ \\
3 & $3512 \times 3512$ pixels, BI & proton & 12.5 MeV & warm & $5 \times 10^9$, $1 \times 10^{10}$, $5 \times 10^{10}$, $1 \times 10^{11}$ protons/cm$^2$ \\
4 & $3512 \times 3512$ pixels, FI & proton & 12.5 MeV & cold & $5 \times 10^9$, $1 \times 10^{10}$, $2 \times 10^{10}$ protons/cm$^2$ \\
5 & $1700 \times 1836$ pixels, FI & proton & 12.5 MeV & warm & $5 \times 10^9$ protons/cm$^2$\\
6 & $1700 \times 1836$ pixels, FI & proton & 12.5 MeV & warm & $1 \times 10^{10}$ protons/cm$^2$ \\
7 & $1700 \times 1836$ pixels, FI & proton & 12.5 MeV & warm & $5 \times 10^{10}$ protons/cm$^2$ \\
8 & $1700 \times 1836$ pixels, FI & proton & 12.5 MeV & warm & $1 \times 10^{11}$ protons/cm$^2$ \\
9 & $3512 \times 3512$ pixels, FI & electron & 0.1 - 1.0 MeV & cold & 1.2 krad \\
\hline
\end{tabular}
\end{table*}

\begin{figure}[!h]
\centering
\includegraphics[width=3.45in]{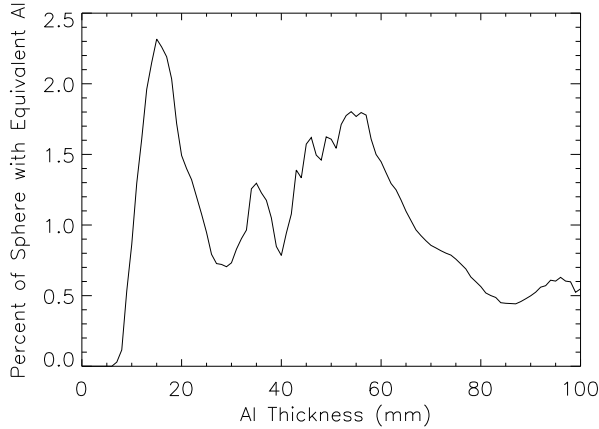}
\caption{ 
Distribution of the shield thickness surrounding the SNAP focal plane over the full $4\pi$ solid angle.
} 
\label{fig:snapshield}
\end{figure}

Analysis of the satellite mechanical structure shows the detector shielding thickness varies by almost a full order of magnitude over the full range of angles of incidence.
The distribution of the material surrounding the focal plane over $4\pi$ is shown
in Figure~\ref{fig:snapshield}. 
The present satellite design provides an average shielding
equivalent to about 47~mm of Al shielding around the focal plane, with a
minimum of 9~mm of Al equivalent over a small fraction of
the solid angle.  The SNAP satellite has not yet been fully optimized for radiation
shielding, and future modifications can provide additional shielding in the thinnest
regions, so our current estimates may be considered conservative.

We have computed the average NIEL at the SNAP focal plane by folding the expected 
NIEL at L2 as a function of shield thickness 
with the distribution of shielding thickness in the current SNAP design.
We find an
integrated NIEL  dose of $6.6 \times 10^6$~MeV/g~(Si).
Assuming a NIEL damage factor of $8.9 \times 10^{-3}$~Mev/g/cm$^2$ for 
$12.5$~MeV protons\cite{nielref}, this is equivalent to a dose of
$7.4 \times 10^8$ 12.5~MeV~protons/cm$^2$.
We report results of the radiation tolerance of the SNAP CCDs treating
this dose as a ``nominal" value that will be experienced by the SNAP CCDs at
$95\%$ CL after six years at L2.

\section{Irradiation at the LBNL 88-Inch Cyclotron} \label{section:irrad}

Nine CCDs were characterized before irradiation, with performance very similar to that
described in Table \ref{table:specs}.
Charge transfer efficiency (CTE) was measured using the $^{55}$Fe 5.9 keV line \cite{bebek} for both
parallel and serial transfers.
Gain conversion from ADC count (ADU) to e$^-$ was also determined using $^{55}$Fe images.
Dark current was determined from the mean signal
in 10 minute dark exposures, after removal of 3$\sigma$ outliers to account
for cosmic ray contamination.
Ten dark images were taken successively and median-combined to generate
a high signal-to-noise dark image, free of cosmic rays and terrestrial background radiation.
Residual hot pixels
caused by a clustering of mid-level traps
were identified as high significance peaks in this median-combined image.
Very rarely was even a single individual hot pixel identified in a dark image
at 133~K. More common were manufacturing defects, the occasional hot column caused by a minor clock short or back-side defect.
For a more detailed account of clock shorts and back-side defects, see \cite{steve2}.

            

To simulate radiation exposure in the space environment,  CCDs 1-8 listed in
Table \ref{table:devices} were exposed at the
LBNL 88-Inch Cyclotron for irradiation to 12.5 and 55~MeV protons.
For convenience, most of the radiation exposures were carried out at room temperature  
on CCDs with all of the inputs shorted together and no bias voltages present.   
The proton fluence was continuously monitored during irradiation using standard
ion chamber dosimetry.

To check whether warm irradiation gives the same results as irradiation at cryogenic temperatures,
a full-size SNAP CCD was irradiated in a dewar at 133~K at
nominal bias and clocking voltages and continuous readout at 70~kHz
during the exposure.
A brass shield inside the dewar could be moved into three different positions, resulting in exposures to three different regions of the CCD.
The cold-irradiated CCDs allowed us to study the time evolution of the dark current,
and the rate at which hot pixels were generated.
In the warm-irradiated devices, both dark current and hot pixels quickly
annealed at room temperature, so only the cold-irradiated CCDs could give an indication of the
long-term effects.  
In addition, we carried out controlled periods of
warming on the cold-irradiated devices to study the effects of annealing.

\begin{figure*}[!b]
\centerline{\subfigure[Parallel CTE as a function of dose]{\includegraphics[width=2.5in,angle=270]{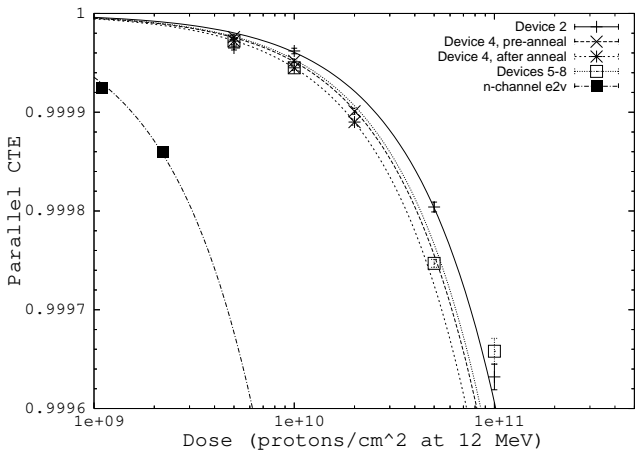}
\label{figure:parallel CTE}}
\subfigure[Serial CTE as a function of dose]{\includegraphics[width=2.5in,angle=270]{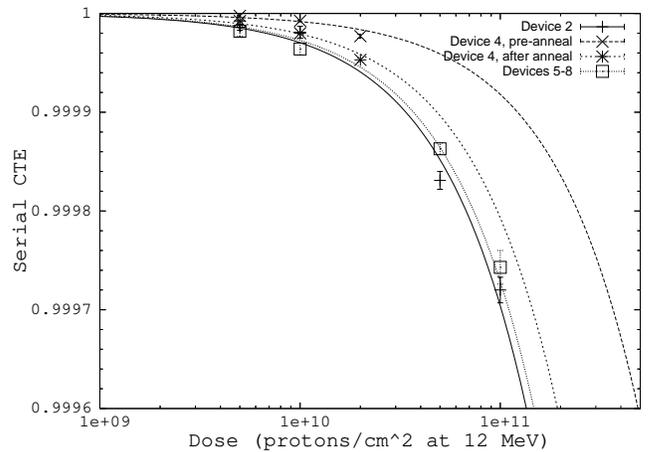}
\label{figure:serial CTE}}}
\caption{a)  Parallel CTE as a function of dose for SNAP CCDs and n-channel e2v CCD similar
to that used in ACS on HST with model fits.
b)  Serial CTE as a function of dose for SNAP CCDs with model fits.
}
\label{fig:CTE}
\end{figure*}

\section{Results of Proton Irradiation} \label{section:results}

Measurements on the warm-irradiated devices were made beginning
four weeks after irradiation to 
allow the dark current to decay to a low level.   Otherwise, the
abnormally high dark current would mitigate the effects of degraded CTE by
filling the defects created during irradiation.
After this cooling off period, the CCDs were
again characterized as described above 
to determine of the CTE as a function of dose and energy.

For CTE measurements, the $^{55}$Fe linear density was approximately
one x-ray per 80 pixels for devices 1-3 and devices 5-8.
The density was approximately one x-ray per 270 pixels for devices 4 and 9.
All CTE measurements were carried out at a temperature of 133~K
at a pixel readout rate of 70~kHz.
Because of the delay between parallel transfers as each row is serially read one pixel at a time,
charge is transfered about three orders of magnitude faster in the serial (line)  direction than in the direction of parallel (row) transfer.
The traps are most efficient when the transfer rate is comparable
to the de-trapping time constant.

The cold, proton-irradiated device 4
was maintained at 133~K for
seven weeks following irradiation.
Dark and $^{55}$Fe images were collected on a regular basis,
beginning three days after the irradiation.
The primary purpose of the cold-irradiation and analysis
was to determine the evolution of CTE,
dark current, and isolated hot pixels at normal operating conditions
over an extended period.
After seven weeks, the device was allowed to anneal to room temperature
for a period of 12 hours and then cooled back down to 133~K for
CTE and dark current measurements.
Measurements were again taken daily for another seven weeks.
Comparison of the CCD performance before and after warming provide
data on the effects of annealing, an analysis not possible
with the warm-irradiated CCDs.

\subsection{Comparison of CTE on front- and back-illuminated CCDs} \label{subsection:bivsfi}

Most of the irradiated devices were 650~$\mu$m thick, front-illuminated (FI)
CCDs.  Front-illumination refers to the
light impinging on the front, or patterned, side of the CCD (CCDs used for astronomy
are always back-illuminated for improved quantum efficiency).
The FI devices lend themselves to CTE
testing since $^{55}$Fe
x-rays  are deposited directly on the pixels, without the lateral charge diffusion that occurs in
back-illuminated (BI) devices. 
One 200~$\mu$m thick, BI SNAP device (device 3) was irradiated
for comparison.  CTE was measured on the irradiated BI  device using the
extended pixel edge response (EPER) and first pixel response (FPR) techniques \cite{janesick}, 
instead of x-rays.  A detailed comparison of the BI device 3 with FI device 2
with  EPER and FPR  showed a similar
degradation of CTE with dose.   
From this we conclude that the use of 650~$\mu$m thick, 
front-illuminated devices for the study of CTE degradation with dose is a reasonable
substitution for 200~$\mu$m thick, back-illuminated devices.

\subsection{Energy Dependence of CTE Degradation} \label{subsection:ctevsenergy}

To test  the validity of the NIEL scaling for CTE degradation, a SNAP
CCD irradiated at 55~MeV  was compared to a SNAP CCD irradiated at 12.5~MeV (devices 1 and 2 in Table~\ref{table:devices}).
As can be seen in Table~\ref{table:energy}, the damage factor describing
serial charge transfer inefficiency (CTI$ = 1-$CTE)
is nearly identical for both energies, well within the uncertainty of the measurement.
The damage factor was observed to be $15\%$ larger in parallel CTE
in the case of the 55~MeV irradiation, a relatively minor difference of $1.5 \, \sigma$.

\begin{table}[h!]
\caption{CTE degradation at 12.5~MeV and 55~MeV for a dose of $1 \times 10^{11}$~protons/cm$^2$.}
\label{table:energy}
\centering
\begin{tabular}{|l||l||c||c||c|}
Energy & Transfer & CTI & NIEL & Damage Factor \\
(MeV) & Direction & $\times 10^{-4}$ & MeV/g (Si) & CTI/Dose/NIEL \\
$ $ & $ $ & $ $ & $\times 10^{-3}$ & $\times 10^{-13}$ \\
\hline
12.5 & parallel & $3.9 \pm 0.3$ & $8.9 $ & $4.4 \pm 0.3 $ \\
55 & parallel & $2.1 \pm 0.2$ & $4.1 $ & $5.1 \pm 0.5 $ \\
12.5 & serial  & $3.1 \pm 0.4$ & $8.9 $ & $3.5 \pm 0.4 $ \\
55 & serial & $1.5 \pm 0.2$ & $4.1 $ & $3.7 \pm 0.5 $ \\
 
\hline
\end{tabular}
\end{table}

\subsection{Scaling of CTE with Dose} \label{subsection:ctevsdose}

\begin{figure*}[!t]
\centerline{\subfigure[Parallel trails]{\includegraphics[width=3.5in,angle=0]{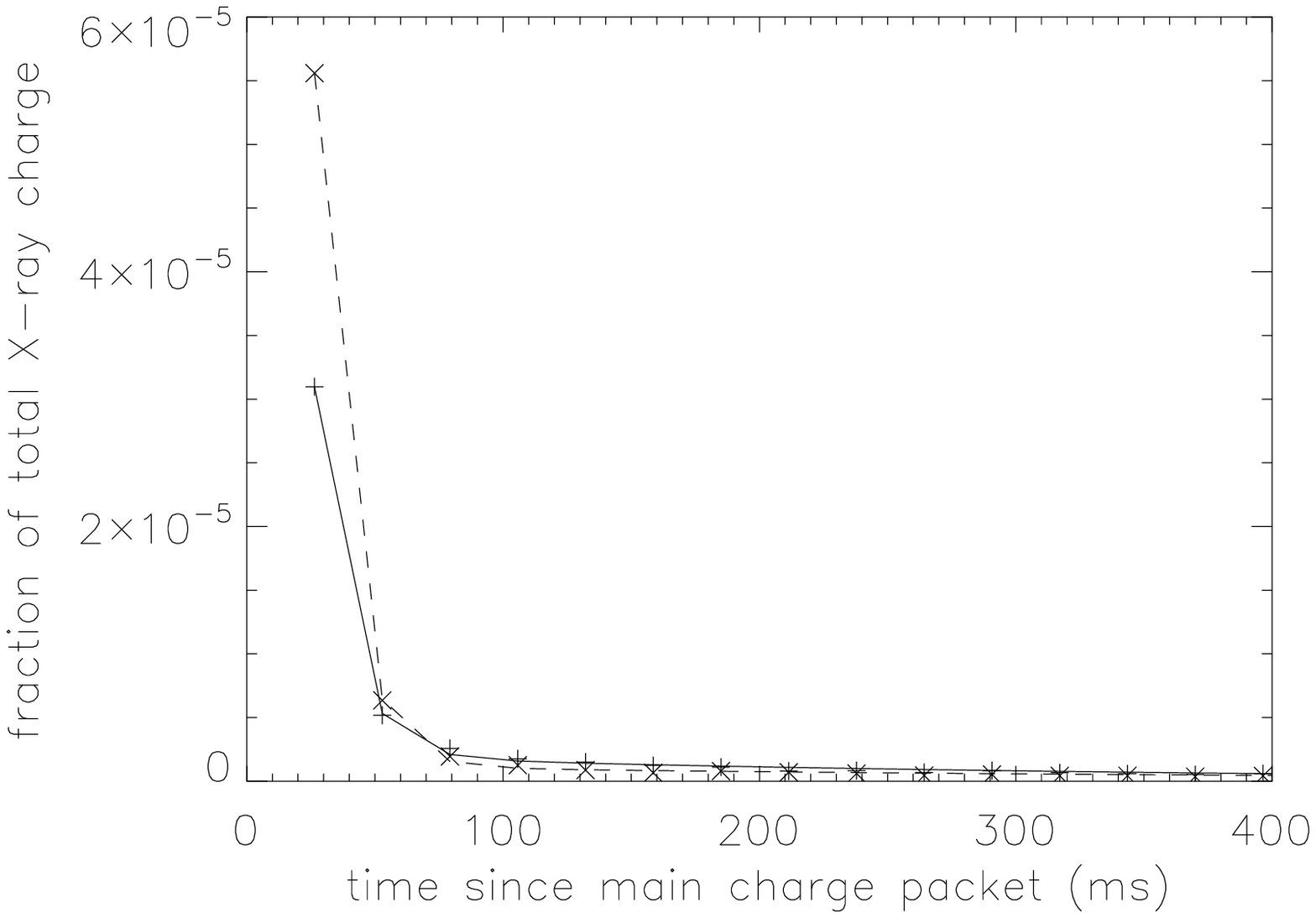}
\label{figure:parallel trails}}
\subfigure[Serial trails]{\includegraphics[width=3.5in,angle=0]{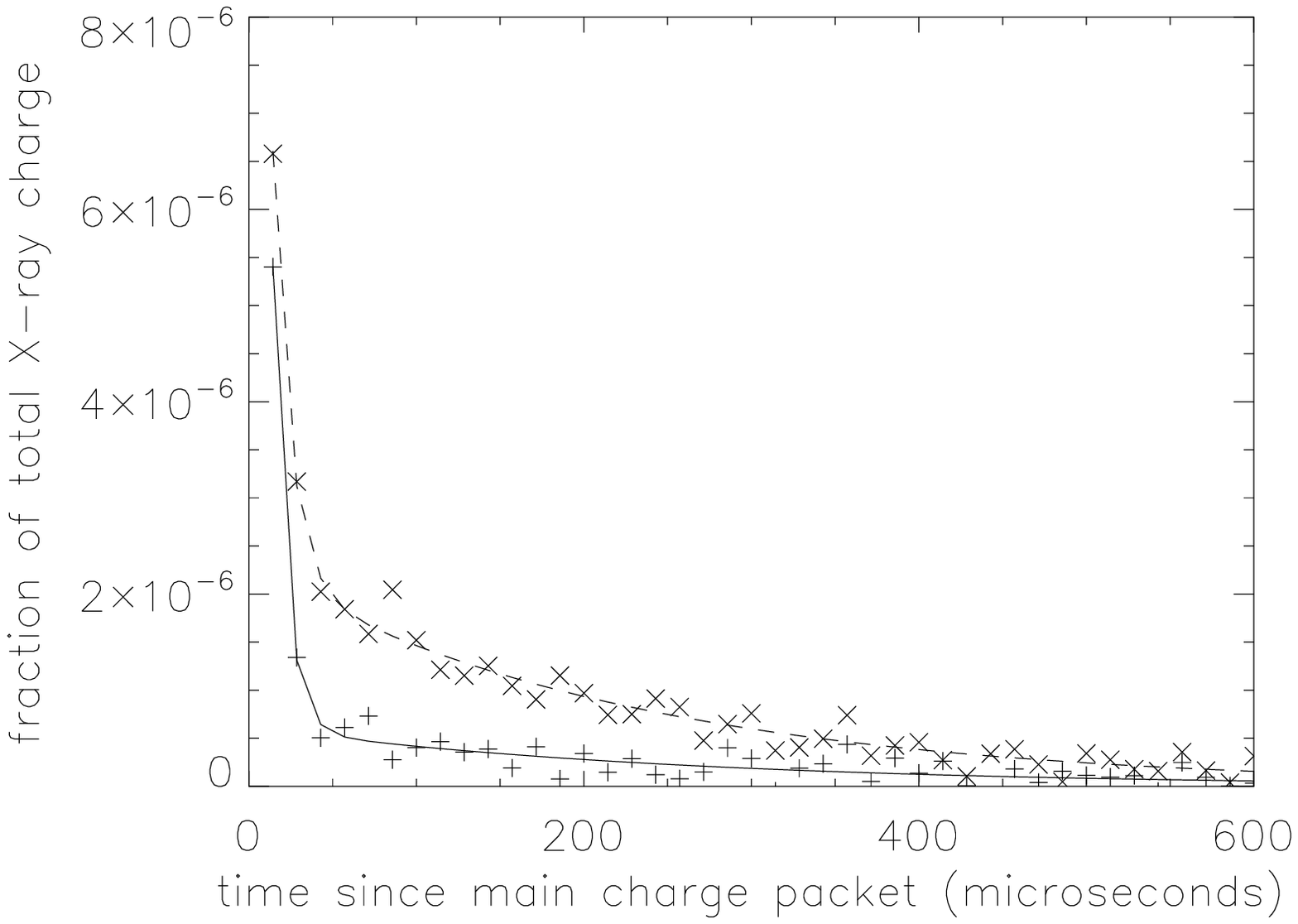}
\label{figure:serial trails}}}
\caption{a)  Distribution of trailing charge in the parallel direction.
Counts are normalized to the number of counts measured in lead pixel, divided by the
number of parallel transfers.
b)  Distribution of trailing charge in the serial direction.
Counts are normalized to the number of counts measured in the lead pixel divided by the
number of serial transfers.
In both cases, the solid curve and '+' symbols
represent results before the anneal.
The dashed curve and 'x' symbol represent results following the anneal.
A data point is taken for every pixel following the main charge.
}
\label{fig:CTE trails}
\end{figure*}

\begin{table*}[!t]
\renewcommand{\arraystretch}{1.0}
\caption{Characterization of Trailing Charge}
\label{table:trails}
\centering
\begin{tabular}{|l||c||c||c||c|}
Measurement & $A_1$ & $\tau_1$ (s) & $A_2$ & $\tau_2$ (s) \\
\hline
Parallel - pre-anneal  & $(1.75 \pm 0.04) \times 10^{-4}$ & $(1.40 \pm 0.02) \times 10^{-2}$ & $(1.84 \pm 0.06) \times 10^{-6}$ & $(4.06 \pm 0.20) \times 10^{-1}$  \\
Parallel - post-anneal & $(5.02 \pm 0.06) \times 10^{-4}$ & $(1.16 \pm 0.01) \times 10^{-2}$ & $(1.28 \pm 0.04) \times 10^{-6}$ & $(3.70 \pm 0.14) \times 10^{-1}$  \\
Serial - pre-anneal  & $(3.42 \pm 0.38) \times 10^{-5}$ & $(7.72 \pm 0.44 ) \times 10^{-6}$ & $(0.64 \pm 0.06) \times 10^{-6}$ & $(2.23 \pm 0.26) \times 10^{-4}$  \\
Serial - post-anneal  & $(2.21 \pm 0.14) \times 10^{-5}$ & $(9.30 \pm 0.34) \times 10^{-6}$ & $(2.22 \pm 0.04) \times 10^{-6}$ & $(2.25 \pm 0.05) \times 10^{-4}$  \\
\hline
\end{tabular}
\end{table*}

The irradiated devices included both full-size $3512 \times 3512$ pixel SNAP CCDs and
"mini-SNAP" CCDs of smaller format $1700 \times 1836$ pixels but of otherwise
identical design.
With the use of a brass shield, the four quadrants of the full-size SNAP CCDs
were individually exposed
to doses of $5 \times 10^9$, $1 \times 10^{10}$, $5 \times 10^{10}$ and $1 \times 10^{11}$
protons/cm$^2$; the mini-SNAP devices each received a single uniform dose.
Comparison of the results for device 2 with devices 5, 6, 7, and 8 (Figures~\ref{figure:parallel CTE}
and \ref{figure:serial CTE}) indicates that the radiation damage
effects observed on the mini-SNAP CCDs were consistent with those observed on
the full-size SNAP CCDs, thus validating the use of small-format devices of
otherwise identical design for radiation studies.

The CTE of devices 2, 4, and 5-8 was analyzed and compared over the full range of exposure levels.
Results of the degradation of parallel CTE are shown in Figure~\ref{figure:parallel CTE}.
There is a slight difference in the parallel CTE among the different
radiation exposure conditions.
This may be due to differences in the level of dark current,
which can account for changes on the order of a few $\times \, 10^{-5}$ in
CTE at a dose of $2 \,\times \, 10^{10}$~protons/cm$^2$, as discussed in \S \ref{subsection:traps}.
The background from dark current in the cold-irradiated device
before the anneal was typically $\sim 10-40$~e$^-$/pix, while the background in the warm-irradiated
devices and in device 4 after annealing was typically $\sim 2-8$~e$^-$/pix.

For comparison, we also include the results of CTE testing on conventional
n-channel CCDs from e2v \cite{marshall} in Figure~\ref{figure:parallel CTE}.
The n-channel CCDs are intended to be used in the Wide Field Camera 3 (WFC3) on the Hubble Space Telescope (HST)
and were irradiated using 63~MeV protons with a fluence of $2.5 \times 10^{9}$~protons/cm$^2$
and $5 \times 10^{9}$~protons/cm$^2$, equivalent to 2.5 and 5.0 years in the HST orbit.
Assuming a NIEL of $3.7 \times 10^{-3}$~MeV/g~(Si) for 63~MeV protons\cite{nielref}, the equivalent dose
at 12.5~MeV is $1.04 \times 10^{9}$~protons/cm$^2$ and $2.08 \times 10^{9}$~protons/cm$^2$.

Serial CTE vs dose is shown in Figure~\ref{figure:serial CTE}.
As can be seen in the figure, the warm-up to room temperature resulted 
in a decrease in the serial CTE, an effect referred to as "reverse annealing."
We also observe a significantly worse  serial CTE performance in 
the warm-irradiated CCDs, compared to the cold-irradiated device both
before and after annealing.
It has been demonstrated that irradiation produces only negligible
degradation of serial CTE in the n-channel e2v devices \cite{hstperf}
and results are not included here.

\subsection{Effect of Annealing on CTE} \label{subsection:anneal}

\begin{table*}[b!]
\renewcommand{\arraystretch}{1.0}
\caption{Best fit parameters to trap-filling model}
\label{tab:trap_params}
\centering
\begin{tabular}{|l||c||c||c||c||c|}
Measurement & $A_1$ & $s_0$ (e$^-$/pixel) & $A_2$ & $\rho_0$ (events/pixel) & C \\
\hline
Parallel & $(5.6 \pm 1.3) \times 10^{-5}$ & $14.6 \pm 3.1$ & $(4.1 \pm 0.6) \times 10^{-5}$ & $(1.2 \pm 0.4) \times 10^{-2}$ & $(5.0 \pm 1) \times 10^{-5}$ \\
Serial & $(2.7 \pm 1.7) \times 10^{-6}$ & $53 \pm 99$ & $(1.7 \pm 1.2) \times 10^{-5}$ & $(1.6 \pm 2.2) \times 10^{-2}$ & $(9.0 \pm 14) \times 10^{-6}$ \\
\hline
\end{tabular}
\end{table*}

Reverse annealing has also been observed in the n-channel
CCDs used in the Chandra telescope.
Following that analysis \cite{chandra}, 
we analyze the de-trapping time constants before and after
annealing by computing the average signal in 
the pixels following the main charge packet in the
$^{55}$Fe images from the cold, proton-irradiated device 4.

Each x-ray event is identified, centroided in $3 \, \times \, 3$ pixel box, and included in the analysis
if the center position is within $0.1$ pixels of the center pixel.
This selection rejects events in which the x-ray is deposited near a pixel
boundary.
The charge is counted in each trailing pixel as a fraction of the charge
in the primary charge packet for the parallel
or serial directions.
We then divide the trail of charge of each event by the total number of transfers
and average the results.
In other words, the averaged trails represent the fraction of charge left behind
the primary charge packet for a single transfer.
The results before and after the anneal for parallel and serial clocking
are found in Figure~\ref{fig:CTE trails}.

The trailing charge is well fit by a two term exponential of the form
\begin{equation}
Q(t) = A_1 \, e^{-t/\tau_1} \, + \, A_2 \, e^{-t/\tau_2}
\end{equation}
where $Q(t)$ is the number of counts following the main charge packet
as a function of time.
The best fits are plotted in Figure~\ref{fig:CTE trails}, and the parameters
are reported in Table~\ref{table:trails}.

One can compute the amount of charge described by both terms
of the exponential decay 
by simply integrating the best fit curve to infinity.
The ratio of the integrals 
\begin{equation}
R = \frac{\int_0^\infty{A_1 \, e^{-t/\tau_1} \, dt}}{\int_0^\infty{(A_1 \, e^{-t/\tau_1} \,+\, A_2 \, e^{-t/\tau_2})\,dt}}
\end{equation}
determines the fraction of charge that is contained in the fast decay
decay term compared to the total charge contained in the trails.

For the parallel CTE, most of the trailing charge is contained in
the fast decay term: $77\%$ before the anneal, and $92\%$
after the anneal.
For the serial CTE, however, a significant difference is observed between
the pre-anneal trailing charge and the post-anneal trailing charge.
Before the anneal, $65\%$ of the trailing charge
is contained in the fast decay term.
After the anneal, the longer decay term dominates, with only
$29\%$ of the charge being contained in the fast decay term.

The significant change in the characteristics of the serial trailing charge
indicates a transition in the trap population caused by the anneal.
Previous studies indicate that divacancies are the traps primarily
responsible for CTE degradation in LBNL CCDs, with carbon interstitials
and carbon-oxygen traps playing a less significant role\cite{bebek}.
It is possible that a population of relatively benign lattice vacancies is
generated during the initial cold irradiation, and remains stable at
low operating temperatures.
If this is the case, it appears that this population
becomes mobile at room temperature,
possibly forming more stable, and more efficient
divacancy traps during the annealing process.
A full diagnosis of the effects of the reverse anneal requires
measurements of pocket-pumping\cite{janesick} and
CTE as a function of temperature to constrain the trap properties
before and after the anneal.
Such an analysis is beyond the scope of
this paper and will be addressed in future publications.

\subsection{Effects of Trap-Filling on CTE Performance} \label{subsection:traps}

\begin{figure*}[!t]
\includegraphics[width=6.05in, angle=0]{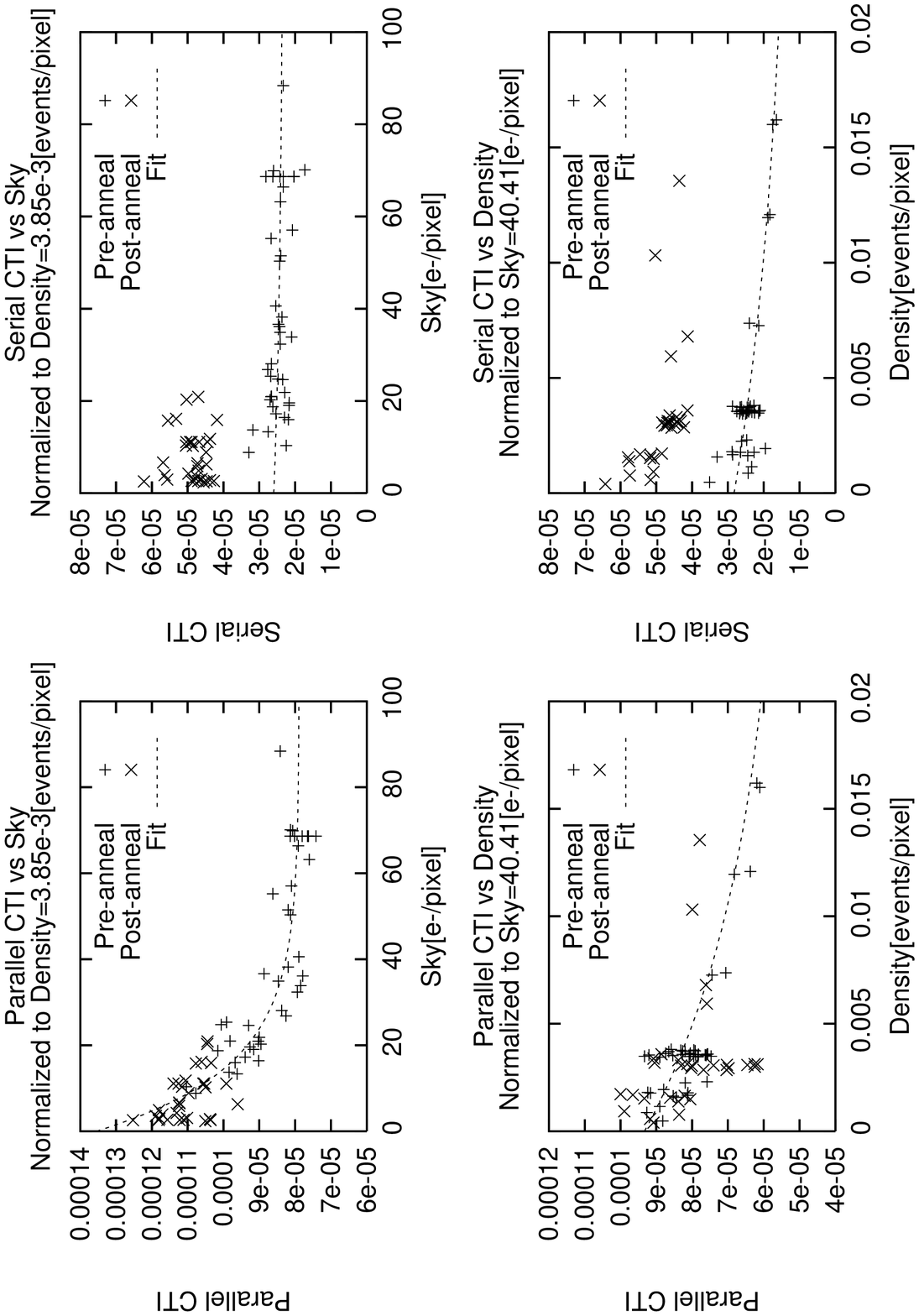}
\caption{
Dependence of CTI on sky level and density of x-ray events for both the parallel and
serial CTI after a dose of $2 \times 10^{10}$~protons/cm$^2$.
Upper left: Parallel CTI vs sky level.
Upper right: Serial CTI vs sky level.
CTI is normalized to x-ray density of
$3.85 \times 10^{-3}$~events/pixel. 
Lower left: Parallel CTI vs x-ray density.
Lower right: Serial CTI vs x-ray density.
CTI is normalized to sky level of
40.41~e$^-$/pixel.
}
\label{fig:traps}
\end{figure*}

It is well known that trap-filling by background sky and neighboring objects can
mitigate the effects of CTE degradation \cite{janesick}.
In this section we describe an effort to model the
dependence of CTE in SNAP CCDs on the background
sky level and the density of $^{55}$Fe events.

Device 4 was imaged with varying exposure times
to control the level of dark current
and varying shutter times to control the density of $^{55}$Fe events.
We took several sets of data,
covering a factor of 40 in both the range of background sky values and
$^{55}$Fe densities, both before and after annealing.

Sky dependent corrections to CTI
have been modeled for observations with the Advanced Camera for Surveys
(ACS)\cite{riess}.
In the corrections to account for trap-filling on the ACS on HST, it 
was assumed that the CTI dependence on both the sky background and
the source intensity is described by a power law.
Such an assumption produces a singularity in the limit of low sky
background or low source intensity. 
The data is quite noisy in both the ACS analysis and in this analysis,
and it is difficult to determine which analytic function best describes the data.
We avoid the singularities introduced by a power law and simply assume
that sky level and source density affect the CTI independently.
We fit the data with an exponential law of the form
\begin{equation}
CTI(s,\rho) = A_1 e^{-s/s_0} + A_2 e^{-\rho/\rho_0} + C
\end{equation}
where $s$ represents the sky level in units of e$^-$/pixel,
$\rho$ is the density of x-ray events in units of events/pixel, and $A_1$,
$A_2$, $s_0$, $\rho_0$, and $C$ are the parameters to be fit.
Parameters are determined by a fit
to the pre-anneal data in the quadrant
that received a radiation dose of $2 \times 10^{10}$~protons/cm$^2$.
The dark current in the post-anneal data was very low and the data were not sufficient
to constrain the model.

The best fit parameters that describe the CTI as a function of background
level and x-ray density are found in Table~\ref{tab:trap_params}.
The CTI data as a function of sky level and density are reported along
with the best fit model in Figure~\ref{fig:traps}.
In the two upper tiles of the figure, CTI is plotted versus sky level after normalization
to $3.85 \times 10^{-3}$ x-ray events per pixel using the best fit parameters.
In the two lower tiles, CTI is plotted versus x-ray density after normalization
to a sky level of 40.41~e$^-$/pixel.
For parallel CTI, both the pre-anneal and post-anneal data are well described
by the same set of parameters.
It is evident from the figure that the serial CTI after the anneal follows a
significantly different relationship than the pre-anneal data, another indication
of a transition in the trap population caused by the anneal cycle.
Also demonstrated in Figure~\ref{fig:traps} is that
the mitigation of CTE from the background
sky and x-ray density is more pronounced in the parallel transfer direction than
in the serial transfer direction.
The biggest improvement appears to come from an increased sky
background, decreasing the parallel CTI from $1.3 \times 10^{-4}$ at
zero background to $8.0 \times 10^{-5}$ at a background of $40-100$~e$^-$/pixel
at the fixed x-ray density of $3.85 \times 10^{-3}$ events per pixel.

\begin{table*}[b!]
\caption{Parameters Describing Evolution of Dark Current}
\label{table:DC}
\centering
\begin{tabular}{|l||c||c||c||c||c|}
Dose & $A_0$ (e$^-$/px/hr) & $t_0$ (hr) & $A_1$ (e$^-$/px/hr) & $t_1$ (hr) & C (e$^-$/px/hr) \\
\hline
\multicolumn{6}{|l|} {Before Room Temperature Anneal} \\
$5\times 10^9$ & 6500 $\pm 40.7$ & 61.9 $\pm 0.7$ & 1050 $\pm 27.8$ & 331 $\pm 9.5$ & 113 $\pm 4.1$ \\
$1\times 10^{10}$ & 12900 $\pm 136$ & 63.4 $\pm 1.2$ & 2100 $\pm 108$ & 328 $\pm 19.5$ & 228 $\pm 20.0$ \\
$2\times 10^{10}$ & 24300 $\pm 470$ & 61.5 $\pm 1.3$ & 4200 $\pm 156$ & 311 $\pm 12.1$ & 466 $\pm 20.5$ \\
$^{60}$Co & (3.4 $\pm 0.24) \times 10^6$  & 0.32 $\pm 0.04$ & (1.0 $\pm 0.08) \times 10^6$ & 17.2 $\pm 1.6$ & (8.8 $\pm 2.7) \times 10^3$ \\
\hline
\multicolumn{6}{|l|} {Following Room Temperature Anneal} \\
$1\times 10^{10}$ & 398 $\pm 90$ & 52.8 $\pm 15$ & 142 $\pm 47$ & 194 $\pm 37$ & 58 $\pm 3$ \\
$2\times 10^{10}$ & 730 $\pm 44$ & 59.6 $\pm 4.9$ & 178 $\pm 21$ & 288 $\pm 32$ & 94 $\pm 2.5$ \\
\end{tabular}
\end{table*}

\subsection{Generation of Hot Pixels} \label{subsection:hotpixels}

\begin{figure}[!h] 
\centering 
\includegraphics[width=2.4in, angle=270]{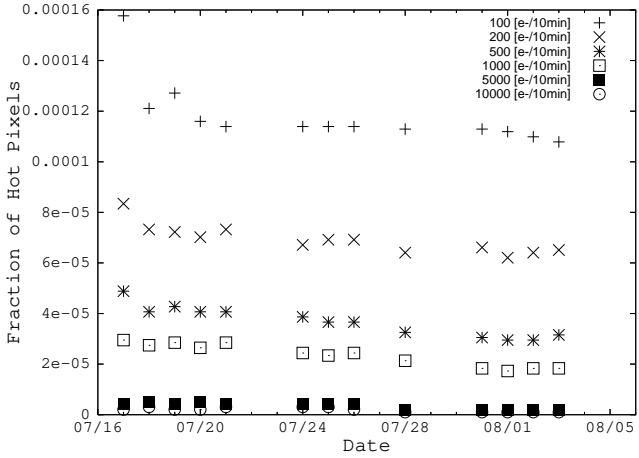}
\caption{Isolated hot pixels after irradiation with $2 \times 10^{10}$~protons/cm$^2$.}
\label{figure:hp}
\end{figure}

Median-stacked, cosmic ray-cleaned dark images from before and after irradiation were compared
in the quadrant which received a dose of $2 \times 10^{10}$~protons/cm$^2$
in the cold-irradiated SNAP CCD (device 4).
Using a simple scheme to subtract the pre-irradiation image from the post-irradiation
image, a map was generated to identify residuals produced as a result of the irradiation.
Hot isolated pixels in this residual map represent spikes in dark current, created from
clustering of bulk defects and will be flagged in a bad
pixel map for science images.
Hot pixels are located and counted by identifying pixels that lie a certain threshold
above the mean background level.
The pre-anneal number density of these hot pixels as a function of time
and threshold is shown in Figure~\ref{figure:hp}.

With a threshold of 100~e$^-$ in a ten minute exposure, the density of hot pixels is $1.13 \times 10^{-4}$
for a dose of $2 \times 10^{10}$~protons/cm$^2$.
The density of hot pixels is $3.1 \times 10^{-5}$ with a threshold of 500~e$^-$ in a ten minute exposure.
After the anneal, the already negligible
density of hot pixels drops dramatically.

A similar experiment was conducted using n-channel CCDs designed by e2v for WFC3.
These CCDs were exposed to 63~MeV protons at a total fluence of $2.5 \times 10^{9}$~protons/cm$^2$,
equivalent to an exposure at 12.5~MeV of $1.04 \times 10^{9}$~protons/cm$^2$.
After the anneal, a fraction of $2.5 \times 10^{-3}$ hot pixels were detected
at a threshold of 26~e$^-$/10~min\cite{polidan}.
Applying this threshold to the LBNL data, and scaling the result to the same dose,
we find a fraction of $2.0 \times 10^{-5}$ hot pixels
before the anneal and $1.3 \times 10^{-6}$ hot pixels after the anneal in the SNAP device.

The improvement by over two orders of magnitude
in the rate of hot pixels for the LBNL CCDs relative to
the e2v CCDs is at least in part due to the different
operating temperatures for the SNAP (-133~C) and WFC3 (-83~C)
focal planes.
The rate of hot pixels in the e2v
CCDs was observed to decline by two orders of magnitude
as operating temperature was reduced from -65~C to -90~C.
The hot pixel rate in LBNL CCDs has not been studied
at the higher temperature of the WFC3 instrument.

\subsection{Evolution of Dark Current} \label{subsection:darkcurrent}

\begin{figure}[h]
\centering
\includegraphics[width=2.4in,angle=270]{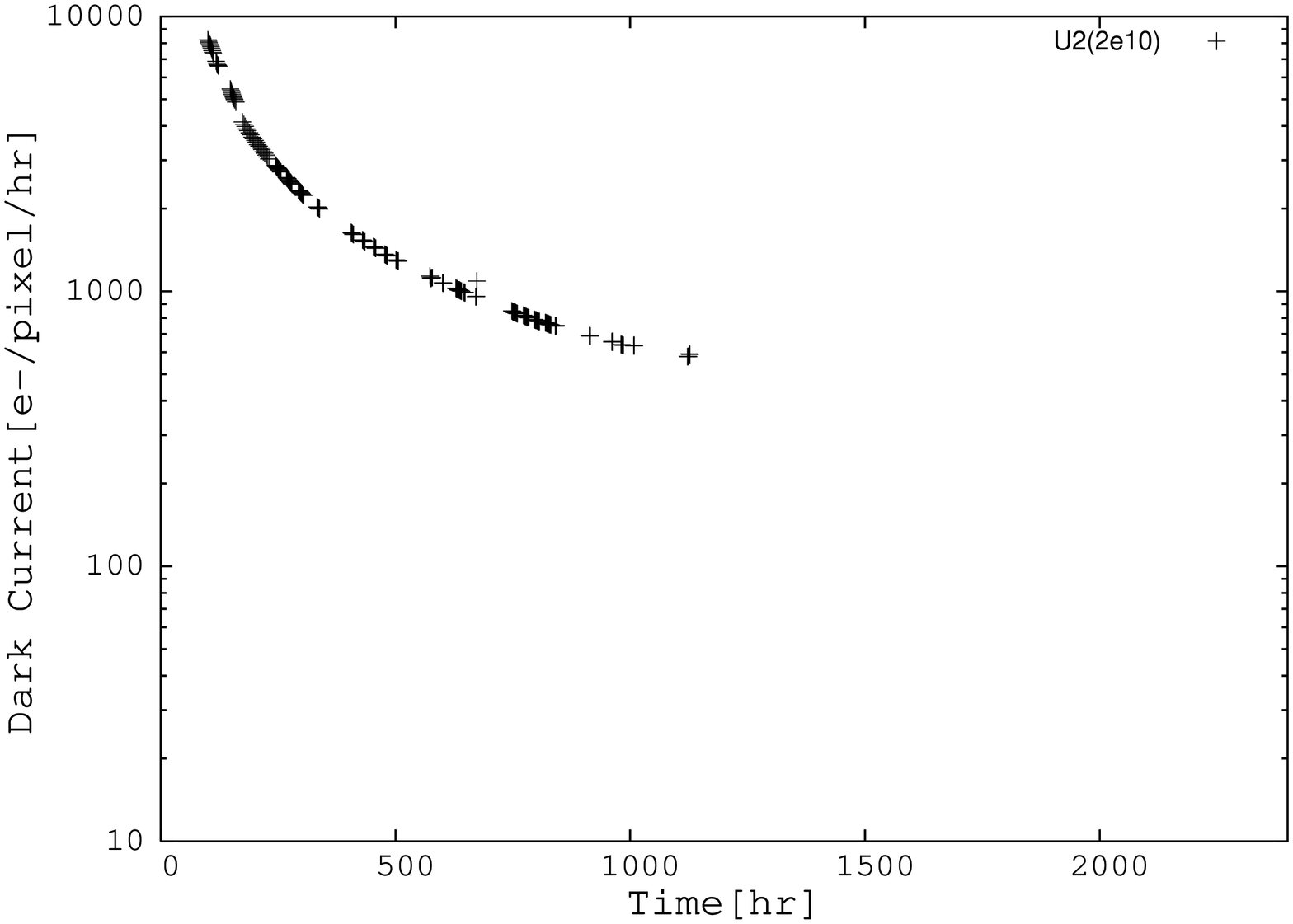}
\caption{Evolution of dark current in cold-irradiated device.}
\label{figure:dark current}
\end{figure}

The level of dark current ($DC$) for device 4 as a function of time can be found in Figure~\ref{figure:dark current}.
The evolution of dark current is well described by a two term exponential decay,
\begin{equation}
DC = A_0\,e^{-t/t_0} \, + \, A_1\,e^{-t/t_1} \, + \, C
\end{equation}
where $A_0$ and $A_1$ describe the amplitude of the two exponential terms, in units
of e$^-$/pixel/hr, and $t_0$ and $t_1$ are the corresponding time constants.
The model is fit to the data, and best-fit parameters can be found in Table~\ref{table:DC}.
The curve described by the best-fit model for each dose is found in Figure~\ref{figure:dark current}.
Examination of the best fit parameters indicate that the dark current scales
roughly with dose before the anneal and that the time constants are not dose dependent.
It is also evident from Table~\ref{table:DC} that the decay time constants are short
compared to the mission lifetime.
A room temperature anneal appears to initiate a second decay in the dark current
with time constants similar to those observed immediately following the exposure.


\section{Irradiation with the LBNL $^{60}$Co Source} \label{section:Co60}

In order to separate the effects of ionizing radiation damage from the effects of
NIEL radiation damage, another SNAP CCD (device 9 in Table~\ref{table:devices} was irradiated at the $^{60}$Co source at LBNL.
The CCD was mounted in a dewar with an Al window of thickness 0.75 mm in
place of the usual glass window.
2 mm of Pb shielding was placed 
in between the dewar and the $^{60}$Co source.
The device was powered and irradiated for 30 minutes 
at a temperature of 133~K.

The primary mechanism for radiation damage in this experiment is energy
deposition from ionizing electrons in the ~0.1 - 1~MeV range.
Electrons are excited from the Pb shielding through Compton scattering of
1.1 and 1.3~MeV $^{60}$Co photons.
The Al window at the dewar opening was designed to prevent the generation of
excess electron-hole pairs from remaining low energy photons. 
An estimate of the
total ionizing dose of 1.2~krad was determined through Monte Carlo simulations
of the propagation of photons and electrons through the Pb
and Al shielding.
The estimate of the Monte Carlo simulations was confirmed within $10\%$
using thermoluminescent
dosimeters (TLDs) placed at various locations between the CCD and the $^{60}$Co source.
After irradiation, measurements of dark current were obtained
for comparison to the cold-irradiated device described
in \S\ref{subsection:darkcurrent}.
The $^{60}$Co-irradiated device was maintained at 133~K for five days
following irradiation.
Dark images were collected several times a day,
starting 30 minutes after irradiation.
No measurable degradation of CTE was observed in the CCD irradiated
at the $^{60}$Co source.


\section{Results of $^{60}$Co Irradiation} \label{section:Co60 results}

\begin{figure}[h!]
\centering
\includegraphics[width=2.4in,angle=270]{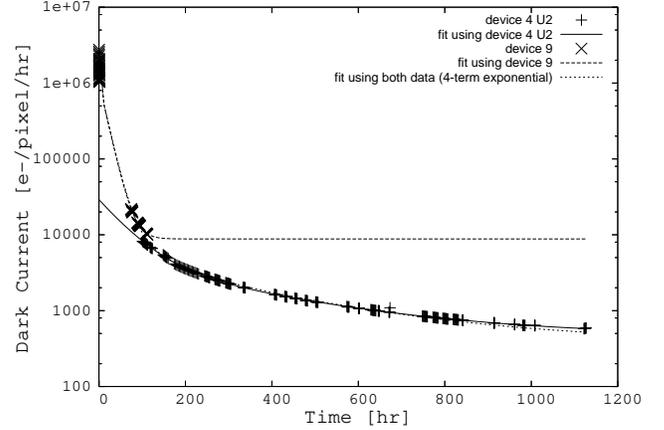}
\caption{Evolution of dark durrent in proton irradiated CCD (device 4)
compared to $^{60}$Co irradiated CCD (device 9)}
\label{figure:Co DC}
\end{figure}

We observed the time evolution of dark current in the $^{60}$Co-irradatiated device.
Assuming the damage scales linearly with dose, the dark current measured in this device
was scaled to a dose of 9.38~krad, the
same ionizing dose experienced by the most damaged quadrant of the proton-irradiated
device.
The results compared to the proton-irradiated CCD
are found in Figure~\ref{figure:Co DC} and
the best fit parameters of the two term exponential model are found in
Table \ref{table:DC}.

The time frame of the measurements for the $^{60}$Co radiation
is quite different from that of the proton-irradiation measurements, and 
the two experiments probe different regions of the dark current decay.
Nevertheless, a comparison of the decay amplitudes
found in Figure~\ref{figure:Co DC} do suggest that the dark current
evolution from ionizing radiation is quite similar to the evolution of dark current
in proton-radiation.
Using the time constants from the two term exponentials for both cases, we find
the combined data is fairly well described by a four term exponential of the form
\begin{equation}
DC = A_1\,e^{-t/0.32} + A_2\,e^{-t/17.2} + A_3\,e^{-t/61.5} + A_4\,e^{-t/311} + C
\end{equation}
where $A_1=(3.43 \pm 0.001) \times 10^6$, $A_2=(9.8 \pm 0.1) \times 10^5$,
$A_3= (2.0 \pm 1.3) \times 10^4$, $A_4=(4.8 \pm 4.1) \times 10^3$,
and $C=(0.4 \pm 1.1) \times 10^3$ in units of e$^-$/pixel/hr.
The curve describing this model can also be found in Figure~\ref{figure:Co DC}.
Although the longer time constants are poorly constrained, the 
values agree quite well with the values from the two term fit to
the proton-irradiated data.

The similarity between the proton-irradiated device
and the $^{60}$Co-irradiated device indicates that ionizing radiation
may be primarily responsible for the generation of dark current.
Similar results have been observed in experiments with other
p-channel devices\cite{spratt}.
One possible explanation may be that the 0.1 - 1.0~MeV electrons occasionally disrupt the lattice,
causing bulk damage and increased dark current without the traps responsible for degraded CTE.
To conclusively determine 
the origins of dark current requires additional irradiation at the $^{60}$Co source
with measurements covering a period of several weeks to provide
constraints on the longer decay constants and  varying the experimental configuration to probe 
the damage caused by electrons and photons of different energies.

\section{Discussion} \label{section:discussion}

Both the parallel and serial CTE scale roughly as expected as a function of proton energy, providing
evidence that the NIEL approximation of CTE degradation is fairly robust.
Assuming the NIEL approximation is valid, we extrapolate the results of the 12.5 MeV irradiation
to model the effects of exposure at the L2 Lagrange point.

\begin{figure}[h!]
\centering
\includegraphics[width=3.45in,angle=0]{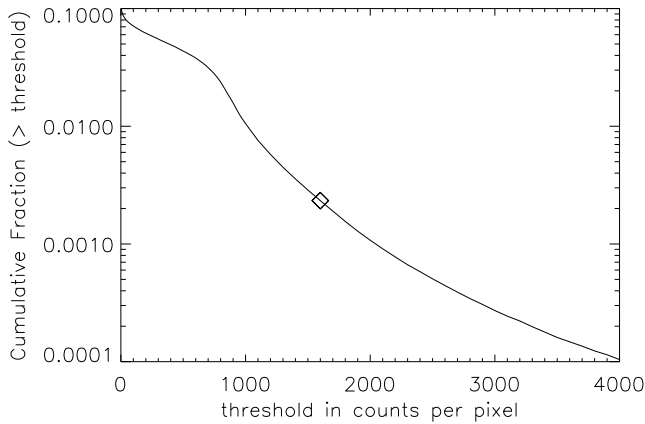}
\caption{Expected cumulative distribution of cosmic ray counts per pixel.
The symbol at 1600~e$^-$ represents the density of cosmic rays above a threshold
corresponding to the average $^{55}$Fe x-ray event.
}
\label{figure:cr density}
\end{figure}

In estimating the performance of the SNAP CCDs after six years at L2, we consider
CTE mitigation by the zodiacal background and cosmic rays in a typical
300 second exposure.
Simulations predict a zodiacal
background of 0.166~e$^-$/s/pixel around 400~nm and 0.446~e$^-$/s/pixel at 1000~nm for the current filter design\cite{davis}.
For the purposes of this analysis, we use the lowest level of
zodiacal from the bluest filter, or $49.8$~e$^-$/pixel for a 300~s SNAP exposure.
We have also computed the expected cosmic ray contamination for a single
SNAP exposure, as shown in Figure~\ref{figure:cr density}.
As a rough estimate, we assume that
trap-filling by $^{55}$Fe x-rays is a fair estimate of trap-filling by cosmic rays
above a threshold of 1600~e$^-$.
We therefore use a value of $2.33 \times 10^{-3}$~events/pixel 
for determining CTE performance in orbit.
Although the results shown in Figure~\ref{figure:cr density} indicate
that a fairly large number of pixels will be contaminated by cosmic rays,
it should be noted that the dithering strategy
will provide multiple exposures that will be used to reject most cosmic rays.

Extrapolating to the expected dose
and modeling the mitigation of CTE according to Equation 3,
the SNAP CCDs are expected to perform extremely well.
Assuming a baseline CTE of $0.999\,999$ before launch,
we calculate a parallel CTE of $0.999\,996$ after six years.
The serial CTE is somewhat better, with a predicted value of $0.999\,997$ after six years. If the device is never annealed, half the serial CTE degradation will occur.

As argued in \S\ref{subsection:hotpixels}, the SNAP CCDs are quite resilient to hot pixels
after irradiation.
Hot pixels effect a very small area of the SNAP CCD, only $1.13 \times 10^{-4}$ for a dose
of $2 \times 10^{10}$~protons/cm$^2$ assuming a threshold of 100~e$^-$ in a ten minute exposure.
Scaling this result to the dose expected, we expect 
$4.1 \times 10^{-6}$ of the pixels to be contaminated by dark current spikes in orbit at L2.
Considering the $3512 \times 3512$ layout of the SNAP CCDs, this level of contamination
is equivalent to a single column defect only 48 pixels long.
The SNAP observing strategy implements a dither pattern to cover gaps between detectors,
equivalent to several hundred columns in width.
The contribution from both column defects and hot pixels will be minor relative to the
spacing between detectors, and the dither pattern will be sufficient to cover any detector
area lost due to these defects.

Finally, we interpret the level of dark current following irradiation in the
context of the SNAP mission.
Ideally, the dominant background in SNAP observations will come from the sky itself,
with the dark current generation in the CCDs playing only a minor role.
We estimate the expected level of dark current after six years
by taking the constant term without annealing, and scaling the dose to the predicted levels from SPENVIS.
After six years with no anneal, the dark current of 20~e$^-$/hr
is significantly lower than the minimum level of zodiacal of 600~e$^-$/hr around 400~nm.
Assuming Poisson statistics, this level of dark current will only increase the RMS contribution
from the background by $2\%$ for the bluest filter.
The situation improves after an anneal.
Dark current due to radiation exposure is therefore not expected to degrade the sensitivity
of SNAP observations of SNe or weak lensing shear.

\section{Conclusion} \label{section:conclusion}

The behavior of thick, fully depleted, p-channel LBNL CCDs designed
for the SNAP satellite has been investigated using
irradiation at the LBNL 88-Inch Cyclotron and LBNL $^{60}$Co source.
We have performed extensive tests of charge transfer efficiency, generation of dark current, and hot pixel
formation from proton exposure.
A summary of the results scaled to the expected exposure at L2 can be found in Table \ref{tab:rad}.
CTE performance after irradiation is calculated assuming pre-radiation
parallel and serial CTEs of $0.999\, 999$.
The radiation studies show that the LBNL CCDs designed for use in the SNAP satellite
will develop negligible contamination from dark current and hot pixels during
the course of a six year mission.

\begin{table}[h]
\renewcommand{\arraystretch}{1.0}
\caption{Expected CCD Performance after Six Years at L2}
\label{tab:rad}
\centering
\begin{tabular}{|c||c||c|}
Quantity & Pre-irrad  & Nominal Dose \\
\hline
Defect Pixels & $<0.001$  &  $4.1 \times 10^{-6}$ \\
Dark Current & $3-4$ e$^-$/hr &  $20$ e$^-$/hr \\
Ser CTE-no anneal & 0.999 999 & 0.999 998  \\
Ser CTE-w/anneal & 0.999 999 & 0.999 997  \\
Parallel CTE & 0.999 999 & 0.999 996  \\
\hline
\end{tabular}
\end{table}

Monte Carlo simulations of radiation exposure by propagation of solar protons through the complex
shielding of the SNAP satellite will finalize estimates of radiation dose over the mission lifetime.
Additional analysis is required to quantify the impact of
the degraded performance on science observations:
CTE degradation impact on galaxy shapes for weak lensing science goals and
CTE dependence on sources signal strength.
Future studies of the effects of $^{60}$Co irradiation over a longer
time span would be useful in better understanding the mechanism for the generation of dark current.

Nevertheless, the results reported here show that the LBNL CCDs are significantly more radiation
tolerant than n-channel CCDs currently in use in space-based observatories.
This makes the LBNL CCDs an excellent choice for use in future space-based
missions such as SNAP.

\section*{Acknowledgment}

The authors would like to acknowledge the contribution of M. Uslenghiin writing
software used to compute CTE values generated in this work.
In addition, the authors thank Chris Stoughton for his comments on
the manuscript.
This work was sponsored by the United States Department
of Energy under contract No. DE-AC02-05CH11231.




%




\end{document}